\documentclass[10pt,letterpaper,twocolumn,conference]{IEEEtran}
\usepackage[cmex10]{amsmath}
\usepackage{amssymb,amsthm}
\usepackage{url}
\usepackage{multicol}
\usepackage[dvips]{graphicx}
\usepackage{psfrag}

\newcommand{\divi}[2]{\frac{\mathrm{d}#1}{\mathrm{d}#2}}
\newcommand{\rdivi}[2]{\frac{\partial #1}{\partial #2}}

\newcommand{\ep}{\epsilon}

\newcommand{\rmax}{d}
\newcommand{\lmax}{b}
\newcommand{\E}[1]{\bar{#1}}
\newcommand{\epbar}{\tilde{\ep}}
\newcommand{\xbar}{\tilde{x}}

\newtheorem{theorem}{Theorem}

\newcommand{\delt}[2]{\delta^{(#1,#2)}}
\newcommand{\cons}[2]{C_{#1,#2}}
\newcommand{\diff}[1]{\mathrm{d}#1}

\title{
Analytical Solution of Covariance Evolution for Regular LDPC Codes}
\author{\IEEEauthorblockN{
Takayuki Nozaki\IEEEauthorrefmark{1}, 
Kenta Kasai\IEEEauthorrefmark{1}, and
Kohichi Sakaniwa\IEEEauthorrefmark{1}}
\IEEEauthorblockA{\IEEEauthorrefmark{1}
Dept.\ of Communications and Integrated Systems,
Tokyo Institute of Technology\\
Email: \{nozaki, kenta, sakaniwa\}@comm.ss.titech.ac.jp
}
}

\begin{document}
\maketitle
\begin{abstract}
The covariance evolution is 
a system of differential equations with respect to
the covariance of the number of edges 
connecting to the nodes of each residual degree.
Solving the covariance evolution,
we can derive distributions of 
the number of check nodes of residual degree 1,
which helps us to estimate the block error probability
for finite-length LDPC code.
Amraoui et al.\ resorted to numerical computations
to solve the covariance evolution. 
In this paper, 
we give the analytical solution of the covariance evolution.
\end{abstract}

\section{Introduction}
Gallager invented low-density parity-check (LDPC) codes \cite{Gallager}
in 1963.
LDPC codes are linear codes defined by sparse bipartite graphs.
Luby et al.\ introduced the \textit{peeling algorithm} (PA) \cite{Luby,mct} 
for the binary erasure channel (BEC).
PA is an iterative algorithm 
which is defined on Tanner graphs. 
PA and brief propagation (BP) decoder
have the same decoding result.
As PA proceeds, edges and nodes are progressively removed.
The residual graphs consist of nodes and edges 
that are still unknown at each iteration.
The decoding successfully halts
if the graph vanishes.

Amraoui \cite{Amraoui} showed that distributions of the number of 
check nodes of degree one in the residual graph
convergences weakly to a Gaussian as blocklength tends to infinity.
Amraoui also showed 
that block and bit error probability of finite-length LDPC codes
are derived by the average and the variance of the number 
of check nodes of degree one in the residual graph.
The average number of check nodes of degree one in the residual graph
is determined from a system of differential equations,
which
was derived and solved by Luby et al.\ \cite{Luby}.
The variance of the number of check nodes of degree one 
in the residual graph
is also determined from a system of differential equations
called \textit{covariance evolution},
which was derived by Amraoui et al.\ \cite{Amraoui}.
Since analytical solution of covariance evolution
has not been known so far,
we had to resort to numerical computations to 
solve the covariance evolution.

An alternative way to determine the variance of the number of
check nodes of degree one
was proposed in \cite{Amraoui}.
The variance of the number of check nodes of degree one
in the residual graph
can be computed by determining the variance of the number of 
erased messages of BP
for BEC with parameter $\ep^{*}$
where $\ep^{*}$ is the threshold of the ensemble 
under BP decoding.
This method is a valid approximation  for the erasure probability 
close to $\epsilon^{*}$.
Moreover, 
Ezri et al.\ extended to this method to
more general channels \cite{ezri2008}.
However, if we solve the covariance evolution analytically,
we can derive the variance of the number of check nodes of degree one
in the residual graph for all $\ep$
where $\ep$ is the channel parameter for the BEC.

In this paper,
we show an analytical solution of the covariance evolution
for regular LDPC code ensembles.

\section{Covariance Evolution \cite{Amraoui}}
In this section,
we briefly review the covariance evolution and \textit{initial covariance} 
in \cite{Amraoui}.

We consider the transmission over the BEC
with channel erasure probability $\epsilon$
using LDPC codes in a ($\lmax$, $\rmax$)-regular LDPC code ensemble.
Let $t$ denote the iteration round 
and $\xi$ be the total number of edges in the original graph.
We define that
\begin{equation}
 \tau 
  :=
 \frac{t}{\xi}. 
 \label{eq:tau}
\end{equation}
Define a parameter $y$ such that 
$\diff{y}/\diff{\tau} = - 1/(\ep y^{\lmax -1})$
and $y = 1$ when $\tau = 0$.
Let $l_{\lmax,t}$ denote a random variable corresponding to the number of
edges connecting to variable nodes of
degree $\lmax$ in the residual graph at the iteration round $t$.
Let $r_{k,t}$ denote a random variable corresponding to the number of
edges connecting to check nodes of 
degree $k$ in the residual graph at the iteration round $t$.
Those random variables depends on 
the choice of the graph from ($\lmax,\rmax$)-regular LDPC code ensemble,
the channel outputs
and the random choices made by PA.
We define 
\begin{equation}
 \mathcal{D}_{t} 
  :=
 \{l_{\lmax,t}, r_{1,t}, r_{2,t}, \dots, r_{\rmax -1,t}\}. \notag
\end{equation}
To simplify the notation, we drop the subscript $t$.
For $i \in \mathcal{D}_{t}$, we define $\E{i}(y)$ by
\begin{align}
 \E{i}(y) := \frac{\mathbb{E}[i]}{\xi}. \notag
\end{align}
We also define $\delt{i}{j}(y)$ by
the covariance of $i$ and $j$ ($i,j \in \mathcal{D}_{t}$)
divided by the total number of edges in the original graph 
i.e.
\begin{align}
 \delt{i}{j}(y) := \frac{\mathrm{Cov}[i,j]}{\xi}. \notag
\end{align}
In \cite{Amraoui},
Amraoui showed these parameters satisfy 
the following system of differential equations
in the limit of the block length.
This system is referred to as covariance evolution.
\begin{align}
 \divi{\delt{i}{j}(y)}{y}
  =
 -\frac{e(y)}{y}
 &\biggl[
  \sum_{k \in \mathcal{D}} 
      \Bigl(\rdivi{\hat{f}^{(i)} }{\E{k}} \delt{j}{k}
           +\rdivi{\hat{f}^{(j)} }{\E{k}} \delt{i}{k}\Bigr)
    \notag \\
  &+\hat{f}^{(i,j)}(y) 
 \biggr], \label{CE}
\end{align}
where
\begin{align}
 &\rdivi{\hat{f}^{(l_{\lmax})}}{\E{l}_{\lmax}} = 0, 
  \hspace{10mm}
  \rdivi{\hat{f}^{(l_{\lmax})}}{\E{r}_j} = 0,  \notag
\end{align}
\vspace{-5mm}
\begin{align}
 &\rdivi{\hat{f}^{(r_j)}}{\E{l}_{\lmax}} 
   = -j(\lmax-1)\frac{\E{r}_{j+1}-\E{r}_j}{\E{l}_{\lmax}^{2}}, \notag \\
 &\rdivi{\hat{f}^{(r_j)}}{\E{r}_k}
   = j \frac{\lmax-1}{\E{l}_{\lmax}}
  \bigl(I_{\{k=j+1\}} -I_{\{k=j\}}  \bigr), 
  \notag 
\end{align}
\vspace{-5mm}
\begin{align}
 &\rdivi{\hat{f}^{(r_{\rmax-1})}}{\E{l}_{\lmax}} 
  = (\rmax-1)(\lmax-1)
  \frac{\E{l}_{\lmax} + \E{r}_{\rmax-1}-\E{r}_{\rmax}}
       {\E{l}_{\lmax}^{2}}, \notag \\
 &\rdivi{\hat{f}^{(r_{\rmax-1})}}{\E{r}_j} 
  = -(1+I_{\{j=\rmax -1\}})\frac{(\rmax-1)(\lmax-1)}{\E{l}_{\lmax}}, \notag 
\end{align}
\begin{align}
 \hat{f}^{(l_{\lmax} l_{\lmax})}
  &= 0, \hspace{10mm}
 \hat{f}^{(l_{\lmax} r_{k})}
  = 0, \notag \\
 \hat{f}^{(r_{k}r_{j})}
  &= kj\frac{\lmax-1}{\E{l}_{\lmax}}
    \Bigl\{
     -\frac{(\E{r}_{k+1}-\E{r}_{k})(\E{r}_{j+1}-\E{r}_{j})}
           {\E{l}_{\lmax}} \notag \\
  &\hspace{-10mm}
     +I_{\{k=j\}}  (\E{r}_{j+1}+\E{r}_{j})
     -I_{\{k=j+1\}} \E{r}_k
     -I_{\{j=k+1\}} \E{r}_j
 \Bigr\}, \notag
\end{align}
where $x:=\ep y^{\lmax -1}$,~ $\xbar:= 1-x$,~
$\epbar := 1 - \ep$,~
$e(y) = \E{l}_{\lmax} = xy$,~
$\E{r}_j = \binom{\rmax-1}{j-1}x^j\xbar^{\rmax-j}$,~
$\E{r}_1 = x(y-1+\xbar^{\rmax-1})$
and $I_{\{k=s\}}$ is the indicator function which equals to 1 
if $k=s$ 
and 0 otherwise.

The initial conditions of the covariance evolution are given by
initial covariances.
The initial covariances are the covariances of the number of edges of each 
degree at the start of the decoding 
divided by the total number of edges in the original graph.
For $j,k \in \{1,2,\dots,\rmax\}$,
 initial covariances are derived in \cite{Amraoui}, as follows.
\begin{align}
 &\delt{l_{\lmax}}{l_{\lmax}}(1)
  =
 \lmax \ep \epbar, \notag \\
 &\delt{l_{\lmax}}{r_j}(1)
  =
 -\lmax \binom{\rmax-1}{j-1}\ep^{j}\epbar^{\rmax -j}(\rmax\ep -j),
 \notag \\
 &\delt{r_j}{r_k}(1) 
  =
 I_{\{k=j\}}j\binom{\rmax-1}{j-1}\ep^{j}\epbar^{\rmax -j}
  \notag \\
 &~~-\rmax\binom{\rmax-1}{j-1}\binom{\rmax -1}{k-1}\ep^{j+k}\epbar^{2d-j-k}
 \notag \\
 &~~+(\lmax -1)\binom{\rmax -1}{j-1}\binom{\rmax -1}{k-1}
  \ep^{j+k-1}\epbar^{2\rmax -j-k-1} 
 \notag \\
 &~~\cdot(\rmax \ep -j)(\rmax \ep -k)
  . \notag
\end{align}

\section{Analytical Solution of Covariance Evolution}

We show in the following theorem
the analytical solution of the covariance evolution,
for a ($\lmax$,$\rmax$)-regular LDPC code ensemble.
The proof 
\footnote{
Taking the derivative of both sides of 
(\ref{res_l_l}), (\ref{res_l_r}) and (\ref{res_r_r})
with respect to $y$,
we can check that those equations fulfill (\ref{CE}).}
is given in Section \ref{sec:derive}.
\begin{theorem} \label{theorem_cov} \upshape
Let $\tau$ be the normalized iteration round of
PA as defined in (\ref{eq:tau}).
For a ($\lmax$,$\rmax$)-regular LDPC code ensemble 
and $j,k \in \{1,2,\dots,\rmax -1\}$,
in the limit of the code length,
we obtain the following.
\begin{align}
 &\delt{l_{\lmax}}{l_{\lmax}}
  = 
 \lmax \ep \epbar, \label{res_l_l} \\
 &\delt{l_{\lmax}}{r_{j}}
  = 
 -G_{j} \{\ep\epbar(\lmax -1) y^{-1} + \epbar x \}
 + I_{\{j=1\}} \lmax \ep\epbar,
  \label{res_l_r}  \\
 &\delt{r_{k}}{r_{j}}
  =
 \frac{\lmax -1}{\lmax} G_{k} G_{j}
 \{\ep\epbar(\lmax -1)y^{-2} - (\ep -\epbar)x y^{-1} + x^{2}\}
  \notag \\
 &-\rmax \binom{\rmax -1}{k-1}\binom{\rmax -1}{j-1}
  x^{k+j} \xbar^{2\rmax -k-j}
  \notag \\
 &+I_{\{k=j\}} \binom{\rmax-1}{k-1} k x^{k} \xbar^{\rmax -k} 
  +I_{\{k=1,j=1\}}(\lmax \ep \epbar -x \xbar)
  \notag \\
 &-\bigl(I_{\{k=1\}}G_{j}+I_{\{j=1\}}G_{k}\bigr)
   \{\ep\epbar(\lmax -1)y^{-1} -\ep x +x^{2}\},
 \label{res_r_r} 
\end{align}
where 
$G_{j} 
 := \binom{\rmax-1}{j-1}x^{j-1}\xbar^{\rmax -j -1}
(\rmax x -j)+ I_{\{j=1\}}$ and
$y$ is defined by $\diff{y}/\diff{\tau} = - 1/(\ep y^{\lmax -1})$
with $y=1$ when $\tau = 0$.
\end{theorem}

\subsection{scaling parameter $\alpha$}
 In \cite{Amraoui}, \textit{scaling parameter} $\alpha$ 
is given by
\begin{align}
 \alpha 
  =
 -\frac{\sqrt{\delt{r_1}{r_1}(\ep^{*},y^{*})}}
      {\sqrt{\xi/n} \rdivi{\E{r}_1}{\ep}\bigr|_{\ep^{*}; y^{*}}},
 \label{alpha}
\end{align}
where $\ep^{*}$ is the threshold of the ensemble under BP decoding
, $y^{*}$ is the non-zero solution of $\E{r}_1(y)$ at the threshold 
, $n$ is the blocklength
and $\xi$ is the total number of edges in the original graph.

We define $x^{*} := \ep^{*}(y^{*})^{\lmax-1}$
and $\xbar^{*} := 1 -x^{*}$.
Since $\E{r}_{1}(\ep^{*},y^{*}) = 0$ and 
$\rdivi{\E{r}_{1}}{y}|_{\ep^{*};y^{*}} =0$,
we see that
$y^{*} = 1 - (\xbar^{*})^{\rmax-1}$ and
$y^{*} = (\lmax -1)(\rmax -1)x^{*}(\xbar^{*})^{\rmax -2}$.
Using those equations, we have from (\ref{res_r_r})
\begin{align}
 \delt{r_1}{r_1}(\ep^{*},y^{*}) 
  =
 \frac{x^{*}y^{*}}{\lmax -1}(y^{*}-x^{*}). \notag  
\end{align}
(Note that $G_1 = \frac{\lmax}{\lmax-1}y^{*}$).
Recall that $\E{r}_{1}(\ep,y) = x(y-1+\xbar^{\rmax-1})$.
We see that
\begin{align}
 \rdivi{\E{r}_1}{\ep}\bigr|_{y^{*}; \ep^{*}}
  =
 -\frac{x^{*}y^{*}}{\ep^{*}(\lmax-1)}.
 \notag 
\end{align}
From (\ref{alpha}), we can obtain
\begin{align}
 \alpha 
  =
 \ep^{*}
  \sqrt{\frac{\lmax-1}{\lmax}
        \bigl(\frac{1}{x^{*}} - \frac{1}{y^{*}} \bigr)}.
 \notag 
\end{align}
This is the same result as in \cite{Amraoui} for regular LDPC code ensembles.

\subsection{Example of Solution of Covariance Evolution}

\begin{figure}[tb]
 \begin{center}
  \includegraphics[width=8cm,clip]{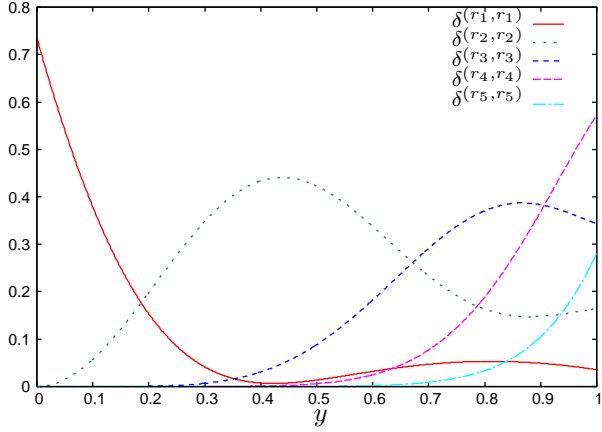}
  \caption{The solution of the covariance evolution $\delt{r_{j}}{r_{j}}$,
  $j\in\{1,2,\dots,5\}$,
  as a function of the parameter $y$ 
  for the (3,6)-regular LDPC code ensemble.
  The channel parameter is $\ep = 0.4294398 \approx \ep^{*} $.}
  \label{fig_cov36}
 \end{center}
\end{figure}

\begin{figure}[tb]
 \begin{center}
  \includegraphics[width=8cm,clip]{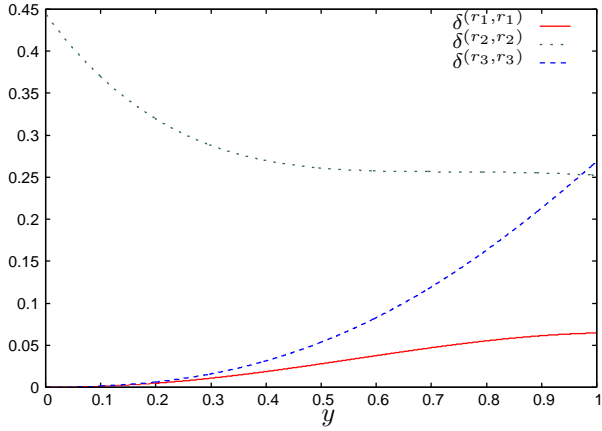}
  \caption{The solution of the covariance evolution $\delt{r_{j}}{r_{j}}$,
  $j\in\{1,2,3\}$,
  as a function of the parameter $y$ 
  for the (2,4)-regular LDPC code ensemble.
  The channel parameter is $\ep = 0.333333 \approx \ep^{*} $.}
  \label{fig_cov24}
 \end{center}
\end{figure}

Figure \ref{fig_cov36} shows 
the solution of the covariance evolution $\delt{r_{j}}{r_{j}}(\ep,y)$, 
$j \in \{1,2,\dots,5\}$,
as a function of $y$ for (3,6)-regular LDPC code ensemble.
Figure \ref{fig_cov24} shows
the solution of the covariance evolution $\delt{r_{j}}{r_{j}}(\ep,y)$, 
$j \in \{1,2,3\}$,
as a function of $y$ for (2,4)-regular LDPC code ensemble.

\psfrag{y}{$y$}
\psfrag{delt11}{\footnotesize{$\delt{r_1}{r_1}$}}
\psfrag{delt22}{\footnotesize{$\delt{r_2}{r_2}$}}
\psfrag{delt33}{\footnotesize{$\delt{r_3}{r_3}$}}
\psfrag{delt44}{\footnotesize{$\delt{r_4}{r_4}$}}
\psfrag{delt55}{\footnotesize{$\delt{r_5}{r_5}$}}

\subsection{Outline of proof} \label{sec:derive}

\subsubsection{Proof for $\delt{l_{\lmax}}{l_{\lmax}}$}

From (\ref{CE}), we get
$\divi{\delt{l_{\lmax}}{l_{\lmax}}}{y}(y)  =  0$. 
From initial covariance, we have
$\delt{l_{\lmax}}{l_{\lmax}} =  \lmax \ep\epbar$.
This leads to (\ref{res_l_l}).

\subsubsection{Proof for $\delt{l_{\lmax}}{r_{j}}$}

In order to solve $\delt{l_{\lmax}}{r_{j}}$,
we define 
$A^{(l_{\lmax},\Sigma)} 
 := \sum_{j=1}^{\rmax-1}\delt{l_{\lmax}}{r_{j}}$,
which gives 
$\divi{A^{(l_{\lmax},\Sigma)}}{y} 
 =  \sum_{j=1}^{\rmax -1}\divi{\delt{l_{\lmax}}{r_{j}}}{y}$.
From (\ref{CE}), we see that
\begin{align}
 \divi{A^{(l_{\lmax},\Sigma)}}{y}
  =
   \frac{\lmax -1}{y} 
  \bigl(\rmax A^{(l_{\lmax},\Sigma)} +  D^{(l_{\lmax},\Sigma)}\bigr),
\end{align}
where
\begin{align}
 D^{(l_{\lmax},\Sigma)}
  :=
 \rmax (x^{\rmax -1}y^{-1} - 1)  \delt{l_{\rmax}}{l_{\rmax}}  .
\notag  
\end{align}
This equation  is a first order linear differential equation
and the solution is given by
\begin{align}
 A^{(l_{\lmax},\Sigma)}
  &=
 y^{\rmax (\lmax-1)}
  \Bigl\{
  \int   
   \frac{(\lmax-1)D^{(l_{\lmax},\Sigma)}}{y^{\rmax (\lmax-1)+1}} \diff{y}
 + \cons{l_{\lmax}}{\Sigma} 
 \Bigr\}
 \notag \\
  &=
   G_{\rmax} \ep\epbar (\lmax-1) y^{-1}  
  + \lmax \ep\epbar 
  + \cons{l_{\lmax}}{\Sigma} y^{\rmax(\lmax-1)} ,
\notag 
\end{align}
with a constant $\cons{l_{\lmax}}{\Sigma}$
determined from the initial covariance
where
$G_{j} 
 := 
 \binom{\rmax -1}{j-1}x^{j-1}\xbar^{\rmax-j-1}(\rmax x-j) +
 I_{\{j=1\}}$.
Note that 
$A^{(l_{\lmax},\Sigma)}(1) = \sum_{j=1}^{\rmax -1}\delt{l_{\lmax}}{r_{j}}(1)
= \lmax \ep\epbar - \lmax \rmax \ep^{\rmax}\epbar $.
We see that
$\cons{l_{\lmax}}{\Sigma} = -\rmax\ep^{\rmax}\epbar$.
We have
\begin{align}
 A^{(l_{\lmax},\Sigma)}
  &=
  G_{\rmax}\{(\lmax-1)\ep\epbar y^{-1} +\epbar x\}
  + \lmax \ep\epbar .
\label{res_A}
\end{align}

From (\ref{CE}) and (\ref{res_A}), 
we get for $j\in\{1,\dots,\rmax-1\}$
\begin{align}
 \divi{\delt{l_{\lmax}}{r_{j}}}{y}
  &=
 \frac{\lmax -1}{y}
 \{ j\delt{l_{\lmax}}{r_{j}}   + D^{(l_{\lmax},r_{j})}  \} ,
\notag
\end{align}
where
\begin{align}
 D^{(l_{\lmax},r_{j})}
  :=&
 j\{\frac{\E{r}_{j+1}-\E{r}_{j}}{\E{l}_{\lmax}}\lmax \ep\epbar
    -\delt{l_{\lmax}}{r_{j+1}}I_{\{j\neq \rmax -1\}} \notag \\
   &+ (A^{(l_{\lmax},\Sigma)} -\lmax \ep\epbar ) I_{\{j =\rmax -1\}}
 \} .
\notag 
\end{align}
Those equations are first order linear differential equations.
The solutions are given by
\begin{align}
 \delt{l_{\lmax}}{r_{j}}
  =
 y^{j(\lmax-1)} \Bigl\{
 \int   \frac{(\lmax-1)D^{(l_{\lmax},r_{j})}}{y^{j(\lmax-1)+1}} \diff{y} 
 +\cons{l_{\lmax}}{r_{j}} 
 \Bigr\},
 \notag
\end{align}
with constants $\cons{l_{\lmax}}{r_{j}}$ 
determined from the initial covariances.
Those equations can be solved by mathmatical induction 
for $j\in\{2,3,\dots,\rmax-1\}$.

We show that $\delt{l_{\lmax}}{r_{\rmax-1}}$ fulfill (\ref{res_l_r}).
From (\ref{res_A}), we can write
\begin{align}
 D^{(l_{\lmax},r_{\rmax-1})}
  =&
  (\rmax-1)\epbar x G_{\rmax} \notag \\
 &+\ep\epbar y^{-1}
  \{\lmax G_{\rmax-1} + (\lmax-1)(\rmax-1)G_{\rmax}\} .  \notag
\end{align}
Using the same way in the induction step,
we can obtain
\begin{align}
 \delt{l_{\lmax}}{r_{\rmax-1}}
  =
 -G_{\rmax-1}\{(\lmax-1)\ep\epbar y^{-1} +\epbar x\}. \notag
\end{align}

We show that if $\delt{l_{\lmax}}{r_{j+1}}$ fulfill (\ref{res_l_r}),
then also $\delt{l_{\lmax}}{r_{j}}$ fulfill (\ref{res_l_r}) .
Assume 
$\delt{l_{\lmax}}{r_{j+1}} = -G_{j+1}\{\ep\epbar(\lmax-1)+\epbar x\}$.
Using the induction hypothesis,
we can write
\begin{align}
 D^{(l_{\lmax},r_{j})}
  =
  j\epbar x G_{j+1}
 +\ep\epbar y^{-1}
  \{\lmax G_{j} + (\lmax-1)jG_{j+1}\} . \label{eq_A_1}
\end{align}
Using $\xbar^{k} = \sum_{s=0}^{k}\binom{k}{s}(-x)^{s}$,
we see that
\begin{align}
 &y^{j(\lmax-1)}\int \frac{(\lmax-1) j\epbar x G_{j+1}}{y^{j(\lmax-1)+1}}
  \diff{y}
  \notag \\
  =&
 -\epbar x G_{j} 
 +\frac{\rmax}{\rmax -j-1} \binom{\rmax-1}{j-1}\epbar x^{j}.
 \label{eq_A_2}
\end{align}
Similarly,
we have
\begin{align}
 &y^{j(\lmax-1)}\int 
 \frac{(\lmax-1) \ep\epbar y^{-1} \lmax G_{j}}{y^{j(\lmax-1)+1}}
  \diff{y}
  \notag \\ 
 =&
 (\lmax-1) \lmax \ep\epbar \binom{\rmax-1}{j-1}x^{j}
  \sum_{s=0}^{\rmax-j} \binom{d-j}{s}(j+s) (-\ep)^{s-1}K_{s-1},
 \label{eq_A_3} \\
 &y^{j(\lmax-1)}\int 
 \frac{(\lmax-1)^{2} \ep\epbar y^{-1} jG_{j+1}}{y^{j(\lmax-1)+1}}
  \diff{y}
  \notag \\  
 =&
 -(\lmax-1)^{2}\ep\epbar \binom{\rmax-1}{j-1}x^{j}
  \sum_{s=0}^{d-j}\binom{d-j}{s}s(j+s)(-\ep)^{s-1}K_{s-1},
 \label{eq_A_4}
\end{align}
where 
\begin{align}
 K_{s}
  :=
 \frac{y^{s(\lmax-1)-1}}{s(\lmax-1)-1}I_{\{s(\lmax-1)\neq 1\}}
 + \log y I_{\{s(\lmax-1) = 1\}}. \notag
\end{align}
Note that
\begin{align}
 \{s(\lmax-1)-\lmax\} K_{s-1}
  =
 y^{(s-1)(\lmax-1)-1}
 -I_{\{(s-1)(\lmax-1)=1\}}. \notag
\end{align}
From (\ref{eq_A_3}) and (\ref{eq_A_4}),
we have
\begin{align}
 &y^{j(\lmax-1)}\int 
 \frac{(\lmax-1) \ep\epbar y^{-1}
  \{\lmax G_{j} + (\lmax-1)jG_{j+1}\}}{y^{j(\lmax-1)+1}}
  \diff{y}  
 \notag \\
 =&
 -(\lmax -1)\ep\epbar \binom{\rmax-1}{j-1}x^{j} \notag \\
 &\hspace{5mm}\cdot\sum_{s=0}^{\rmax-j}\binom{\rmax -j}{s}(j+s)
 \{s(\lmax-1)-\lmax\}K_{s-1}
\notag \\
 =&
 -(\lmax-1)\ep\epbar y^{-1} G_{j} 
 +(\lmax-1)\ep\epbar x^{j} P_{j},
 \label{eq_A_5}
\end{align}
where
\begin{align}
 P_{j}
  :=
 \binom{\rmax-1}{j-1}
 \sum_{s=0}^{\rmax-j} \binom{\rmax-j}{s}(j+s)(-\ep)^{s-1}
  I_{\{(s-1)(\lmax-1)=1\}}. \notag
\end{align}
From (\ref{eq_A_1}),(\ref{eq_A_2}) and (\ref{eq_A_5}),
we have
\begin{align}
 \delt{l_{\lmax}}{r_{j}}
  =&
 -G_{j}\{(\lmax-1)\ep\epbar y^{-1} + \epbar x\}
 + \frac{\rmax}{\rmax-j-1} \binom{\rmax-1}{j-1}\epbar x^{j}
 \notag \\
 &+(\lmax-1)\ep\epbar x^{j} P_{j}
  + \cons{l_{\lmax}}{r_{j}}y^{j(\lmax-1)}. \notag 
\end{align}
From initial covariance,
we have
\begin{align}
 \cons{l_{\lmax}}{r_{j}}
  =
 - \frac{\rmax}{\rmax-j-1} \binom{\rmax-1}{j-1}\epbar \ep^{j}
 -(\lmax-1)\ep\epbar \ep^{j} P_{j} .
 \notag 
\end{align}
Hence we obtain
\begin{align}
 \delt{l_{\lmax}}{r_{j}}
  =
 -G_{j}\{(\lmax-1)\ep\epbar y^{-1} + \epbar x\}. \notag
\end{align}
This lads to (\ref{res_l_r}) for $j\in\{2,3,\dots,\rmax-1\}$.

Note that 
$\delt{l_{\lmax}}{r_{1}} = 
A^{(l_{\lmax},\Sigma)} - \sum_{j=2}^{\rmax -1}\delt{l_{\lmax}}{r_{j}}$
and that
$-G_{1} = \sum_{j=2}^{\rmax} G_{j}$.
We have
\begin{align}
 \delt{l_{\lmax}}{r_{1}}
  =
 -G_{1}\{(\lmax-1)\ep\epbar y^{-1} + \epbar x\}
 + \lmax\ep\epbar .
\notag
\end{align}
Hence we obtain (\ref{res_l_r}).

\subsubsection{Proof for $B^{(\cdot, \Sigma)}$}

In order to solve $\delt{r_{j}}{r_{k}}$ ,
we define
$B^{(r_{j},\Sigma)} 
 :=
 \sum_{k=1}^{\rmax -1}\delt{r_{j}}{r_{k}}$ 
and
$B^{(\Sigma,\Sigma)} 
 :=
 \sum_{j=1}^{\rmax -1}B^{(r_{j},\Sigma)}$.
From (\ref{CE}),
we get for $j\in\{1,2,\dots,\rmax -1\}$
\begin{align}
 \divi{B^{(\Sigma,\Sigma)}}{y}
  =&
 \frac{\lmax -1}{y}
  \bigl( D^{(\Sigma,\Sigma)} + 2\rmax B^{(\Sigma,\Sigma)}  \bigr),
    \label{diff_B}  \\
 \divi{B^{(r_{j},\Sigma)}}{y}
  =&
 \frac{\lmax -1}{y}
 \bigl\{ D^{(r_{j},\Sigma)} + (\rmax + j)B^{(r_{j},\Sigma)} \bigr\},
    \label{diff_Br}
\end{align}
where
\begin{align}
 D^{(\Sigma,\Sigma)}
  :=&
 d\frac{\E{r}_{\rmax} -\E{l}_{\lmax}}{\E{l}_{\lmax}}
 \bigl( \rmax \E{r}_{\rmax}  + 2  A^{(l_{\lmax},\Sigma)} \bigr),
  \notag \\
 D^{(r_{j},\Sigma)}
 :=&
  j \frac{\E{r}_{j+1}-\E{r}_{j}}{\E{l}_{\lmax}} 
  (\rmax \E{r}_{\rmax} + A^{(l_{\lmax}, \Sigma)})
 + \rmax \frac{\E{r}_{\rmax} - \E{l}_{\lmax}}{\E{l}_{\lmax}} 
    \delt{l_{\lmax}}{r_{j}}
  \notag \\
 &-j B^{(r_{j+1},\Sigma)} I_{\{j \neq \rmax -1\}}
  \notag \\
 &+(\rmax -1) (B^{(\Sigma,\Sigma)} -\rmax r_{\rmax} - A^{(l_{\lmax},\Sigma)})
   I_{\{j=\rmax -1\}}.
 \notag 
\end{align}
The solution of (\ref{diff_B}) is given by
\begin{align}
 B^{(\Sigma,\Sigma)}
  =&
  y^{2\rmax(\lmax-1)} \Bigl\{
 \int(\lmax-1) \frac{D^{(\Sigma,\Sigma)}}{y^{2\rmax(\lmax-1)+1}}   \diff{y}
 + \cons{\Sigma}{\Sigma} 
 \Bigr\}
 \notag \\
 =&
 \frac{\lmax -1}{\lmax}G_{\rmax}^{2}
 \{   \ep\epbar (\lmax-1) y^{-2} - (\ep -\epbar)  x y^{-1}  \}
   \notag \\
 &+2G_{\rmax}
 \{ (\lmax-1) \ep\epbar y^{-1} + \epbar  x   \}
   \notag \\
 &+ \rmax x^{\rmax}
  + \lmax\ep\epbar       
  + \cons{\Sigma}{\Sigma} y^{2\rmax(\lmax-1)} ,
 \notag
\end{align}
with a constant $\cons{\Sigma}{\Sigma}$ 
which determined from the inital covariance.
From the initial covariance,
we get 
\begin{align}
 B^{(\Sigma,\Sigma)}(1)
  = 
  \lmax\ep\epbar 
 -2\lmax\rmax\ep^{\rmax}\epbar
 +\rmax\ep^{\rmax}
 -\rmax\ep^{2\rmax}
 +(\lmax-1)\rmax^{2}\ep^{2\rmax -1}\epbar. \notag
\end{align}
We see that
$\cons{\Sigma}{\Sigma}
 = \frac{\lmax -1}{\lmax}\rmax^{2} \ep^{2\rmax} 
 - \rmax\ep^{2\rmax}$.
Hence we have
\begin{align}
 B^{(\Sigma,\Sigma)}
  =&
 \frac{\lmax -1}{\lmax}G_{\rmax}^{2}
  \{(\lmax-1)\ep\epbar y^{-2} -(\ep -\epbar)xy^{-1}+x^{2}\}
   \notag \\
 &+2G_{\rmax}\{(\lmax-1)\ep\epbar y^{-1} +\epbar x\}
   \notag \\
 &+\rmax x^{\rmax} - \rmax x^{2\rmax}
  +\lmax\ep\epbar.
 \label{res_B}
\end{align}

From (\ref{diff_Br}), we get
\begin{align}
 B^{(r_{j},\Sigma)}
  =&
  y^{(\rmax+j)(\lmax-1)} \Bigl\{
 \int \frac{(\lmax-1)D^{(r_{j},\Sigma)}}{y^{(\rmax+j)(\lmax-1)+1}}   \diff{y}
 + \cons{r_{j}}{\Sigma} \Bigr\},
 \notag 
\end{align}
with constants $\cons{r_{j}}{\Sigma}$.
For $j\in\{2,3,\dots,\rmax-1\}$, 
those equation are solved by mathmatical induction
as the proof for $\delt{l_{\lmax}}{r_{j}}$.
From the initial covariances,
note that
\begin{align}
B^{(r_{j},\Sigma)}&(1)
  =
 \rmax (\lmax-1) \binom{\rmax -1}{j-1} 
  \ep^{\rmax+j-1} \epbar^{\rmax-j} (\rmax\ep -j) \notag \\
 &+\rmax \binom{\rmax -1}{j-1} \ep^{\rmax+j} \epbar^{\rmax-j}
  -\lmax \binom{\rmax -1}{j-1} \ep^{j} \epbar^{\rmax-j} (\rmax\ep -j). 
\notag
\end{align}
We have
\begin{align}
 B^{(r_{j},\Sigma)}
  =&
 -\frac{\lmax -1}{\lmax} G_{\rmax} G_{j} 
  \{(\lmax-1)\ep\epbar y^{-2} -(\ep-\epbar)xy^{-1} + x^{2}\}
   \notag \\
 &-G_{j} \{(\lmax-1)\ep\epbar y^{-1}+ \epbar x\}
  +\binom{\rmax -1}{j-1} \rmax x^{\rmax+j} \xbar^{\rmax-j} .
  \label{res_Br}
\end{align}

Using 
$B^{(r_{1},\Sigma)} 
 = B^{(\Sigma,\Sigma)}-\sum_{j=2}^{\rmax -1}B^{(r_{j},\Sigma)}$
, we have
\begin{align}
 B^{(r_{1},\Sigma)}
  =&
 -\frac{\lmax -1}{\lmax} G_{\rmax} G_{1} 
  \{(\lmax-1)\ep\epbar y^{-2} -(\ep-\epbar)xy^{-1} + x^{2}\} 
   \notag \\
 &-G_{1} \{(\lmax-1)\ep\epbar y^{-1}+ \epbar x\}
  + \rmax x^{\rmax+1} \xbar^{\rmax -1} 
   \notag \\
 &+ G_{\rmax}\{(\lmax-1)\ep\epbar y^{-1}+\epbar x\}
         +\rmax x^{\rmax}\xbar +\lmax\ep\epbar .
 \label{res_B1}
\end{align}

\subsubsection{Proof for $\delt{r_{k}}{r_{j}}$}

From (\ref{CE}),
we get for $k,j \in\{1,2,\dots,\rmax -1\}$
\begin{align}
 \divi{\delt{r_{k}}{r_{j}}}{y}
  &=
  \frac{\lmax -1}{y}
  \{(k+j) \delt{r_{k}}{r_{j}} + D^{(r_{k},r_{j})}\},
 \notag
\end{align}
where
\begin{align}
 &D^{(r_{k},r_{j})}
  :=
   H_{k,j}
 + H_{j,k}
 -\frac{\E{l}_{\lmax}}{\lmax -1} \hat{f}^{(r_{k},r_{j})} ,
 \notag \\
 &H_{k,j}
  :=
  k\frac{\E{r}_{k+1}-\E{r}_{k}}{\E{l}_{\lmax}}\delt{l_{\lmax}}{r_{j}} 
 -k \delt{r_{k+1}}{r_{j}} I_{\{k \neq \rmax -1\}}
   \notag \\
 &\hspace{10mm}-(\rmax -1) 
  (\delt{l_{\lmax}}{r_{j}} -B^{(r_{j},\Sigma)}) I_{\{k=\rmax -1\}}.
  \notag 
\end{align}
The solutions of those differential equations are given by
\begin{align}
 \delt{r_{k}}{r_{j}}
  &=
 y^{(k+j)(\lmax-1)} \Bigl\{
 \int \frac{(\lmax-1)D^{(r_{k},r_{j})}}{y^{(k+j)(\lmax-1)+1}}   \diff{y}
 + \cons{r_{k}}{r_{j}}  \Bigr\},
\notag
\end{align}
with constants $\cons{r_{k}}{r_{j}}$.
Using (\ref{res_Br}),
we can solve those equations by mathmatical induction
for $j,k\in\{2,3,\dots,\rmax -1\}$.
We have
\begin{align}
 \delt{r_{k}}{r_{j}}
  =&
 \frac{\lmax -1}{\lmax} G_{k} G_{j}
  \{\ep\epbar(\lmax-1)y^{-2}-(\ep-\epbar)xy^{-1}+x^{2}\}
  \notag \\
 &-d\binom{\rmax -1}{k-1}\binom{\rmax -1}{j-1}x^{k+j} 
  \xbar^{2\rmax -k-j}
  \notag \\
 &+I_{\{k=j\}}\binom{\rmax -1}{j-1}j x^{j} \xbar^{\rmax -j} .
  \notag 
\end{align}
Note that
$\delt{r_{k}}{r_{1}} =
 B^{(r_{k},\Sigma)} - \sum_{j=2}^{\rmax -1}\delt{r_{k}}{r_{j}}$.
We have for $k\in\{2,\dots,\rmax -1\}$ 
\begin{align}
 \delt{r_{k}}{r_{1}}
  =&
 \frac{\lmax -1}{\lmax} G_{k} G_{1}
  \{\ep\epbar(\lmax-1)y^{-2}-(\ep-\epbar)xy^{-1}+x^{2}\}
   \notag \\
 &-\rmax \binom{\rmax -1}{k-1} x^{k+1} \xbar^{2\rmax -k-1}
  \notag \\
 &-G_{k} \{(\lmax-1)\ep\epbar y^{-1}-\ep x +x^{2}\}.
\notag
\end{align}
Since
$\delt{r_{1}}{r_{1}} =
 B^{(r_{1},\Sigma)} - \sum_{j=2}^{\rmax -1}\delt{r_{1}}{r_{j}}$,
we have
\begin{align}
 \delt{r_{1}}{r_{1}}
  =&
 \frac{\lmax -1}{\lmax} G_{1}^{2}
  \{\ep\epbar(\lmax-1)y^{-2}-(\ep-\epbar)xy^{-1}+x^{2}\}
   \notag \\
 &-\rmax x^{2} \xbar^{2\rmax -2}
 + x \xbar^{\rmax -1} 
  \notag \\
 &-2G_{1}   \{(\lmax-1)\ep\epbar y^{-1}-\ep x +x^{2}\}
 +(\lmax\ep\epbar -x\xbar).
\notag
\end{align}
Thus, we can obtain (\ref{res_r_r}).

\section{Relationship To Stability Condition}

In this section,
we consider the relationship 
between the \textit{stability condition} \cite{design,mct} 
and
$\lim_{y\to 0} \delt{l_{\lmax}}{r_{1}}(\ep,y)$.

For a $(\lmax,\rmax)$-regular LDPC code ensemble ($\lmax \ge 3$),
we see from (\ref{res_l_r}) and (\ref{res_r_r}) that
\begin{align}
 &\lim_{y\to 0}\delt{l_{\lmax}}{r_{j}}(\ep,y)
  =
  \lmax \ep \epbar I_{\{j=1\}},
\notag \\
 &\lim_{y\to 0}\delt{r_{j}}{r_{j}}(\ep,y)
  =
  \lmax \ep \epbar I_{\{j=1\}}.
\notag
\end{align}
For a $(2,\rmax)$-regular LDPC code ensemble,
we see from (\ref{res_l_r}) and (\ref{res_r_r}) that 
\begin{align}
 &\lim_{y\to0}\delt{l_{\lmax}}{r_{j}}(\ep,y)
  =
 \begin{cases}
  2\ep\epbar \{1-(\rmax-1)\ep\}, & \text{if} ~~j = 1 \\
  2\ep\epbar (\rmax-1)\ep,       & \text{if} ~~j =2  \\ 
  0,                             & \text{otherwise},
 \end{cases} \notag \\
 \lim_{y\to0}&\delt{r_{j}}{r_{k}}(\ep,y) \notag \\
  =&
 \begin{cases}
  2\ep\epbar\{1-(d-1)\ep\}^{2}, 
     & \text{if} ~~ j=k=1 \\
  2\ep^{2}\epbar(\rmax-1)\{1-(\rmax-1)\ep\},
     & \text{if} ~~ (j,k) = (1,2), (2,1) \\
  2\ep\epbar(\rmax-1)^{2}\ep^{2},  
     & \text{if} ~~ j=k=2 \\
  0 ,
     & \text{otherwise.}
 \end{cases} \notag
\end{align}

If we define the correlation coefficient for $i,j \in \mathcal{D}$ by
\begin{align}
 \rho_{i,j}(\ep) 
  := 
 \lim_{y\to 0}\frac{\delt{i}{j}(\ep,y)}
 {\sqrt{\delt{i}{i}(\ep,y) \delt{j}{j}(\ep,y)}},
\notag
\end{align}
we obtain
\begin{align}
 \rho_{l_{\lmax},r_{1}}(\ep)
  =
 \begin{cases}
  1, & \text{if} ~~ I_{\{\lmax =2\}}(\rmax-1)\ep \le 1 \\
 -1, & \text{if} ~~ I_{\{\lmax =2\}}(\rmax-1)\ep > 1.
 \end{cases} \label{eq_stability}
\end{align}
Note that 
$I_{\{\lmax =2\}}(\rmax-1)\ep \le 1$
agree with the stability condition for regular LDPC code ensembles.

\begin{figure}[tb]
 \begin{center}
  \includegraphics[width=8cm,clip]{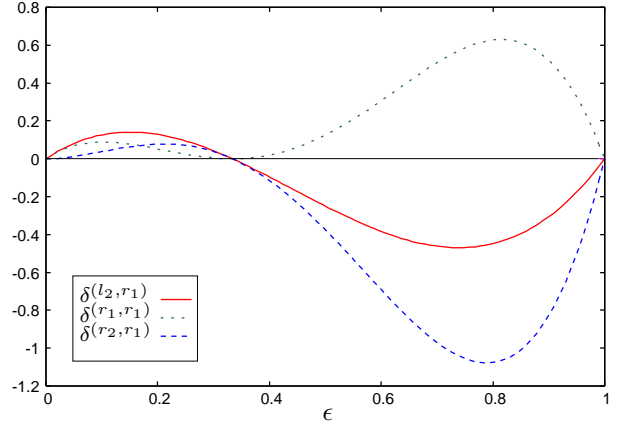}
  \caption{Solution of covariance evolution 
  $\lim_{y\to0}\delt{j}{r_{1}}(\ep,y)$,
  $j\in\{l_{2},r_{1},r_{2}\}$,
  as a function of the channel parameter $\ep$ 
  for the (2,4)-regular LDPC code ensemble.}
  \label{fig_cov24_0}
 \end{center}
\end{figure}

Figure \ref{fig_cov24_0} shows
the solution of covariance evolution $\lim_{y\to0}\delt{j}{r_{1}}(\ep,y)$,
$j\in\{l_{2},r_{1},r_{2}\}$,
as a function of the channel parameter $\ep$ 
for the (2,4)-regular LDPC code ensemble.
From Figure \ref{fig_cov24_0}, 
we see that 
$\delt{r_{2}}{r_{1}} > 0$ and $\delt{l_{2}}{r_{1}} > 0$
when $(d-1)\ep < 1$.
Also we see that
$\delt{r_{2}}{r_{1}} < 0$ and $\delt{l_{2}}{r_{1}} < 0$
when $(d-1)\ep > 1$.

\psfrag{ep}{$\ep$}
\psfrag{delt01}{\footnotesize{$\delt{l_2}{r_1}$}}
\psfrag{delt12}{\footnotesize{$\delt{r_2}{r_1}$}}

\section{Conclusion and Future Work}

In this paper, we have solved analytically
the covariance evolution 
for regular LDPC code ensembles.
Moreover we have derived the relationship between stability condition.

As a future work,
we will derive an analytical solution 
of the covariance evolution for irregular LDPC code ensembles.



\begin{thebibliography}{9}
\bibitem{Gallager}
R.\ G.\ Gallager, 
\textit{Low-Density Parity-check Codes},
MIT Press, 1963.


\bibitem{Luby}
M.\ Luby, M.\ Mitzenmacher, A.\ Shokrollahi, D.\ A.\ Spielman, and V.\ Stemann.
``Practical Loss-Resilient Codes,''
in 
Proceedings of the 29th annual ACM Symposium on Theory of Computing,
1997, pp. 150-159.

\bibitem{Amraoui}
A.\ Amraoui
``Asymptotic and finite-length optimization of LDPC codes,''
Ph.D. Thesis, EPFL, June 2006.

\bibitem{mct}
T.\ Richardson and R.\ Urbanke, \textit{Modern Coding Theory},
Cambridge University Press, 2008.

\bibitem{ezri2008}
J.\ Ezri, A.\ Montanari, S.\ Oh, and R.\ Urbanke.
``The Slope Scaling Parameter for General 
Channels, Decoders, and	Ensembles,''
in Proc.\ of the IEEE Int.\ Symposium on Inform.\ Theory, 
Toronto, 2008.

\bibitem{design}
T.\ J.\ Richardson, M.\ A.\ Shokrollahi, and R.\ L.\ Urbanke, 
``Design of capacity-approaching irregular low-density parity-check codes,''
\textit{IEEE Transactions on Information Theory}, 
vol. 47, no. 2, pp. 619-637, 2001.

\end{thebibliography}
\end{document}